# All-optical neuromorphic binary convolution with a spiking VCSEL neuron for image gradient magnitudes

Yahui Zhang,[1,2] Joshua Robertson,[1] Shuiying Xiang,[2,*] Matěj Hejda,[1] Julián Bueno,[1] and Antonio Hurtado[1,*]

[1] *Institute of Photonics, SUPA Dept. of Physics, University of Strathclyde, TIC Centre, 99 George Street, Glasgow G1 1RD, United Kingdom*
[2] *State Key Laboratory of Integrated Service Networks, Xidian University, Xi'an 710071, China*
[*] *Corresponding author: jxxsy@126.com; antonio.hurtado@strath.ac.uk*

**All-optical binary convolution with a photonic spiking vertical-cavity surface-emitting laser (VCSEL) neuron is proposed and demonstrated experimentally for the first time. Optical inputs, extracted from digital images and temporally encoded using rectangular pulses, are injected in the VCSEL neuron which delivers the convolution result in the number of fast (<100 ps long) spikes fired. Experimental and numerical results show that binary convolution is achieved successfully with a single spiking VCSEL neuron and that all-optical binary convolution can be used to calculate image gradient magnitudes to detect edge features and separate vertical and horizontal components in source images. We also show that this all-optical spiking binary convolution system is robust to noise and can operate with high-resolution images. Additionally, the proposed system offers important advantages such as ultrafast speed, high energy efficiency and simple hardware implementation, highlighting the potentials of spiking photonic VCSEL neurons for high-speed neuromorphic image processing systems and future photonic spiking convolutional neural networks.** © 2020 Chinese Laser Press

## 1. INTRODUCTION

Convolutional neural networks (CNNs) have seen tremendous success in many applications, such as speech and image recognition [1,2], computer vision [3] and document analysis [4]. However, CNN-based systems are computationally expensive due to their complicated architectures and the large number of parameters they rely on. CNNs therefore typically require the implementation of multicore central processing units and graphics processing units to compensate for the rather high computational expense [5-6]. This makes CNN architectures often unsuitable for smaller devices like phones and smart cameras where power and speed have strict limitations. To address these drawbacks, the optimization and the discovery of new high speed and low power consumption platforms for CNNs are urgently required. For the optimization of CNNs, binary CNNs, which are simple, efficient, and accurate approximations of complete CNNs, can be introduced [7-9]. In binary CNNs, the weights given to the inputs of each convolutional layer are approximated with binary values [7]. Therefore, binary CNNs boast 58 × faster convolutional operations and 32 × less memory requirements than those of traditional CNNs [7]. Several optimized binary versions of CNNs have been proposed for training processes and image classification tasks [7,10-11]. However, beyond the optimization of CNNs, a new platform offering high speed and low power consumption remains highly desirable.

Photonics is considered a highly promising candidate for future neural network implementations given the unique advantages it provides such as high speed, wide bandwidth and low power consumption [12–21]. Photonics based CNNs have therefore been proposed in order to increase the speed of convolutional operations [18-21]. A photonic CNN accelerator was proposed based on silicon photonic micro-ring weighting banks [18]. The full system design offers more than 3 orders of magnitude improvement in execution time, and its optical core potentially offers more than 5 order of magnitude improvement compared to state-of-the-art electronic counterparts [18]. S. Xu et al also proposed a high-accuracy optical convolution unit architecture based on acousto-optical modulator arrays where the optical convolution unit was shown to perform well on inferences of typical CNN tasks [20]. However, the size of the system is based on the size of the kernel utilized in these emerging works on photonic CNNs.

In this work, we propose an all-optical binary convolution system using a single vertical-cavity surface-emitting laser (VCSEL) operating as a spiking optical neuron; hence dramatically reducing hardware requirements. In our approach, temporal encoding is used instead of spatial encoding, thus crucially helping to reduce (optical) hardware complexity. In our all-optical binary convolution technique, results are represented by the number of fast (<100 ps long) spiking responses delivered by the optical spiking VCSEL neuron. This has unique advantages in terms of robustness to noise and high precision. Additionally, VCSELs have unique inherent advantages, such as high energy efficiency, high speed modulation capability, low bias currents, easy packaging and highly integrable structures [22,23]. In particular, VCSELs have demonstrated the ability to generate fast spiking dynamics analogous to those of biological neurons known for their robustness to input noise [24-29]. The controlled activation, inhibition and communication of these neuronal dynamics has been demonstrated and recently a single VCSEL device was used to perform spiking pattern recognition and rate-coding [24-31]. Thus, photonic spiking VCSELs make suitable candidates for a new future photonic platform for ultrafast energy efficient spiking CNNs.

In this work, we use a VCSEL-based photonic approach for binary convolution to demonstrate image gradient magnitude calculation. This delivers an essential portion of the image edge detection functionality used by computer vision and image recognition systems. Here, a single VCSEL system is developed to solely perform a convolution operation, hence no VCSEL-based CNN architecture, capable of providing learning and classification capabilities, is discussed in this work. The rest of the paper is organized as follows. Section 2 is devoted to the experimental setup of this work for the demonstration of all-optical binary convolution with a spiking VCSEL neuron and the theoretical model used to predict the response of the system. In Section 3, convolutional results are analyzed before the full calculation of image gradient magnitudes is performed both experimentally and theoretically. Finally, Section 4 summarizes the conclusions of this work.

## 2. EXPERIMENTAL SETUP AND THEORETICAL MODEL

We present here the experimental arrangement and the theoretical model of the all-optical binary convolution system based on a photonic spiking VCSEL neuron. In this work we set a source digital image and a kernel as the two inputs of the convolution system. The value of any one pixel in the source image or kernel is limited to 0 or 1.

### A. EXPERIMENTAL SETUP

Figure 1 shows the schematic diagram of the fiber-optic experimental setup. Two separate electrical signals are generated with a high-bandwidth arbitrary waveform generator (AWG) representing the source image and the kernel used for the convolution process, respectively. These electrical signals (from Channels 1 and 2 of the AWG) are individually amplified by RF Amplifiers 1 and 2, before they are fed into two 10 GHz Mach-Zehnder intensity modulators (Mod1 and Mod2) to encode the source image and kernel into an external optical signal. The latter is generated by a 1300 nm tunable laser (TL). An optical isolator (OI) is included after the TL to avoid unwanted light reflections that might lead to spurious results. A variable optical attenuator (VOA) is used after the OI to adjust the strength of the light signal from the TL. The polarization of the optical signal from the TL is adjusted using three polarization controllers (PC1, PC2 and PC3), where PC1 and PC2 are specifically used to match the polarization of the optical signal to that which maximizes the performance of the two modulators, encoding respectively the image (Mod1) and the kernel (Mod2) information into the optical path. PC3 is used to adjust the final polarization of the encoded optical signal such that it matches the polarization of the targeted VCSEL mode. A 50:50 optical coupler (OC1) is used to split the light signal into two paths. The first one is connected to a power meter (PM) to monitor the input strength, whilst the second one is directly injected into a commercially-available 1300nm-VCSEL through an optical circulator (CIRC). The output of the VCSEL, acting as a spiking optical neuron, is sent to an 8 GHz real-time oscilloscope (SCOPE) and an optical spectrum analyzer (OSA) for analysis. The VCSEL was kept at a constant temperature of 293 K with an applied bias current of 6.5 mA (the lasing threshold current of the VCSEL was $I_{th}$ =2.96 mA at 293 K. The optical spectrum of the free-running VCSEL is shown in Fig. 2(a),

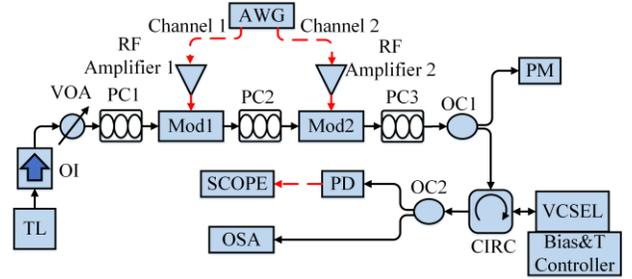

**Fig. 1.** Experimental setup of the binary convolution system based on a single VCSEL. TL: tunable laser; OI: optical isolator; VOA: variable optical attenuator; PC1, PC2, and PC3: polarization controllers; AWG: arbitrary waveform generator; Mod1, Mod2: Mach-Zehnder modulators; OC1, OC2: optical couplers; CIRC: circulator; Bias & T Controller: bias and temperature controller; PD: photodetector; PM: power meter; SCOPE: oscilloscope; OSA: optical spectrum analyzer.

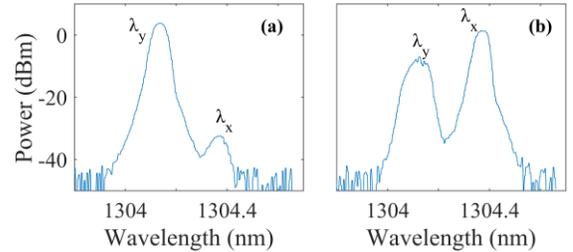

**Fig. 2.** Optical spectra of free-running VCSEL used in the experiment (a). Optical spectra of the VCSEL subject to constant optical injection (b). Two polarization modes of VCSEL are referred as $\lambda_y$ (parallel) and $\lambda_x$ (orthogonal)

where the two lasing peaks correspond to the two orthogonal polarizations of the fundamental transverse mode of the device. We refer to the main lasing mode as the parallel polarized mode (or Y-polarized mode, YP mode, $\lambda_y$) and to the subsidiary mode as the orthogonally polarized mode (or X-polarized mode, XP mode, $\lambda_x$). Figure 2(b) shows in turn the optical spectrum of the 1300nm-VCSEL device in the spiking regime as it is subject to optical injection into the orthogonally polarized mode of the device. Upon injection of the external optical signal into the XP mode of the device, the XP mode becomes the dominant mode whilst the YP mode becomes attenuated. The frequency detuning between the external optically injected signal and the XP mode of the VCSEL was equal to -5.64 GHz. The power of the optically injected signal was 127 µW.

### B. THEORETICAL MODEL

We use an extension of the well-known spin-flip model (SFM) to model the operation of the VCSEL acting as a spiking optical neuron. In our formulation we add extra terms to the model's equations to account for the source image and kernel inputs. The rate equations can be described as follows [26, 27]:

$$\frac{dE_{x,y}}{dt} = -(k \pm \gamma_a)E_{x,y} - i(k\alpha \pm \gamma_p)E_{x,y} + k(1+i\alpha)(NE_{x,y}$$
$$\pm inE_{y,x}) + k_{inj}[E_{injx1}(t) + E_{injx2}(t)]e^{i\Delta\omega_x t} + F_{x,y} \quad (1)$$

$$\frac{dN}{dt} = -\gamma_N[N(1+|E_x|^2+|E_y|^2) - \mu + in(E_y E_x^* - E_x E_y^*)] \quad (2)$$

$$\frac{dn}{dt} = -\gamma_s n - \gamma_N[n(|E_x|^2+|E_y|^2) + iN(E_y E_x^* - E_x E_y^*)] \quad (3)$$

where the subscripts $x$, $y$ represent the XP and YP modes of the VCSEL, respectively. $E_{x,y}$ is the slowly varying complex amplitude of the field in the XP and YP modes. $N$ is the total carrier inversion between conduction and valence bands. $n$ is the difference between carrier inversions with opposite spins. $k$ denotes the field decay rate. $\gamma_a$ and $\gamma_p$ are the linear dichroism and the birefringence rate, respectively. $\alpha$ is the linewidth enhancement factor. $\gamma_N$ is the decay rate of $N$. $\gamma_s$ is the spin-flip rate. $\mu$ represents the normalized pump current. $k_{inj}$ is the injected strength and, $E_{injx1}$ and $E_{injx2}$ indicate respectively the source image and kernel inputs. $\Delta\omega_x$ is defined as $\Delta\omega_x = \omega_{injx} - \omega_0$, where $\omega_{injx}$ is the angular frequency of the externally injected light in the XP mode, $\omega_0 = (\omega_x + \omega_y)/2$ is the center frequency between the XP and YP modes with $\omega_x = \omega_0 + \alpha\gamma_a - \gamma_p$ and $\omega_y = \omega_0 - \alpha\gamma_a + \gamma_p$. The frequency detuning between the externally injected signal and the XP mode is set as: $\Delta f_x = f_{injx} - f_x$. Hence, in Eq. (1), $\Delta\omega_x = 2\pi\Delta f_x + \alpha\gamma_a - \gamma_p$. $F_{x,y}$ are the spontaneous emission noise terms which can be written as:

$$F_x = \sqrt{\frac{\beta_{sp}\gamma_N}{2}}(\sqrt{N+n}\xi_1 + \sqrt{N-n}\xi_2) \quad (4)$$

$$F_y = -i\sqrt{\frac{\beta_{sp}\gamma_N}{2}}(\sqrt{N+n}\xi_1 - \sqrt{N-n}\xi_2) \quad (5)$$

where $\beta_{sp}$ is the strength of the spontaneous emission and, $\xi_1$ and $\xi_2$ are independent complex Gaussian white noise terms of zero mean and a unit variance. We numerically solve Eqs. (1) - (4) using the fourth-order Runge-Kutta method. The parameter values configured for the 1300 nm VCSEL are as follows [26]: $k = 185 ns^{-1}$, $\gamma_a = 2 ns^{-1}$, $\gamma_p = 128 ns^{-1}$, $\alpha = 2$, $\gamma_N = 0.5 ns^{-1}$, $\gamma_s = 110 ns^{-1}$, $\beta_{sp} = 10^{-6}$ and $k_{inj} = 125 ns^{-1}$. With these parameters, the YP mode is the main lasing mode, and the XP mode is subsidiary mode, as in Fig. 2(a).

## 3. EXPERIMENTAL AND NUMERICAL RESULTS

In this section, we firstly provide an experimental proof-of-concept demonstration of all-optical binary convolution with a spiking VCSEL neuron. We then calculate the image gradient magnitudes from a basic "Square" source image and a complex "Horse head" source image by means of all-optical binary convolution. Simulation results on the binary convolution and the calculation of image gradient magnitudes are also presented using a "Horse" source image from the latest version of the Berkeley Segmentation Data Set [32]. Finally, the robustness of our binary convolution system is also tested numerically by adding noise to the source image and kernel inputs.

### A. EXPERIMENTAL RESULTS

Figure 3 shows an example of a binary 2D convolution calculation, where a $3 \times 3$ submatrix (9 pixels) from a source image and a kernel are element-wise multiplied and the subsequent values of the multiplication are summated. In our experiment, we temporally encoded each pixel of the source image and the kernel inputs using rectangular pulses. Pixels of value "1" were optically encoded using intensity modulated power drops in the TL's light (via MZ modulators, Mods 1-2) whereas pixels of value "0" produced no intensity modulation in TL's light. The duration of each rectangular pulse encoding a pixel was set to 1.5 ns to match the refractory period of the experimentally-measured spiking dynamics from the VCSEL neuron [16]. The experimental optical realization of the binary convolution example provided in Fig. 3 is depicted graphically in Fig. 4. Figures 4(a) and 4(b) plot respectively the temporally-encoded 9 pixel ($3 \times 3$) image submatrix and kernel inputs generated for the example given in Fig. 3. Given that the optically-encoded source image and kernel inputs were injected into the VCSEL synchronously, we delayed the kernel input such that its modulation (in Mod2) occurred on-top of the corresponding modulated image input (from Mod 1). We introduced a delay time in the kernel input (directly using the AWG) equal to the time required for a light pulse to travel from Mod1 to Mod2. Figure 4(c) shows the optical signal measured after Mod2 in the setup, combining in a single input line the temporal image and kernel information given in Figs. 4(a) and 4(b). This signal, which was injected into the VCSEL neuron to perform the binary convolution had three different levels (low, medium and high) depending on the specific pixel values in the image and kernel at a given instance. We control the conditions of the injected signal (in Fig. 4(c)) in such a way that the medium and high input levels injection-lock the VCSEL to the external signal, delivering a constant stable temporal output. The lowest input level brings the VCSEL out of the injection-locking and into a dynamical region where the device produces fast spiking dynamical responses [24]. Figure 4(d) shows the experimentally measured time-series at the VCSEL neuron's output, yielding stable or spiking outputs depending on the input intensity levels (from Fig. 4(c)). Importantly, Fig. 4(d) shows that the number of spikes fired by the VCSEL neuron directly provides the result of the binary convolution. It can be seen in Fig. 4(d) that four fast (<100 ps long) spikes are fired by the VCSEL neuron, the same result as that of the binary convolution example in Fig. 3.

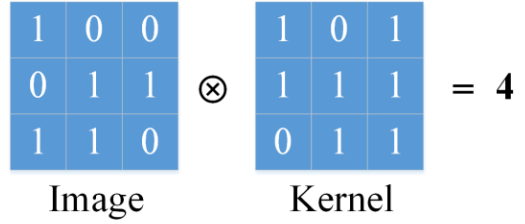

**Fig. 3.** Example of a single step during a 2D binary convolution operation. During this step, a Hadamard (element-wise) product is calculated for a submatrix of the image and the kernel, and all the values in the multiplication result are summed up to obtain a single value.

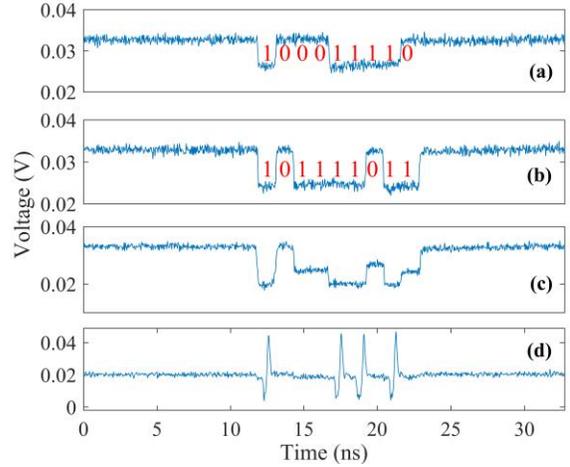

**Fig. 4.** Experimental convolution operation. (a) Inputs of Channel 1 (Image in Fig. 3). (b) Inputs of Channel 2 (kernel in Fig. 3). (c) Inputs of VCSEL. (d) Outputs of VCSEL (the results of convolution).

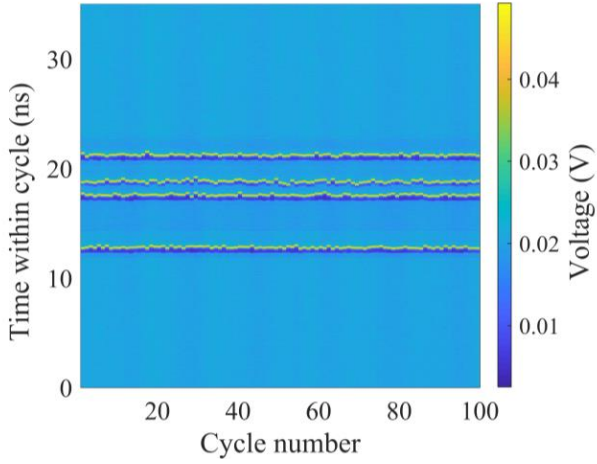

**Fig. 5.** Temporal map of 100 superimposed consecutive convolutional results measured experimentally at the output of spiking VCSEL neuron.

Figure 5 shows a temporal map [17] merging in a single plot 100 superimposed consecutive convolutional outputs from the photonic spiking VCSEL neuron. The image and kernel inputs and the experimental conditions are the same as those shown in Fig. 4. Spike events are depicted in yellow in the colour map of Fig. 5 and steady state responses appear in light blue. Figure 5 clearly shows that binary convolutional result to 100 consecutive inputs remains the same producing, in all 100 cases, 4 separate spiking responses at the VCSEL's output. The optical binary convolutional results obtained with the spiking VCSEL neuron are therefore consistent and reproducible. This proof-of-concept result obtained with a spiking VCSEL highlights a new, controllable way to perform convolution operations for information and image processing tasks.

### B. CALCULATION OF IMAGE GRADIENT MAGNITUDES

In this section, image gradient magnitude, critical to image edge detection, is calculated using our approach based on a single spiking VCSEL neuron and optical binary convolution. The image gradient magnitude $G(x)$ of a given pixel $x$ is calculated using the following equations [33]:

$$G(x) = \sqrt{G_X(x)^2 + G_Y(x)^2} \tag{6}$$

$$G_X(x) = (B(x) \otimes B_X^+) - (B(x) \otimes B_X^-) \tag{7}$$

$$G_Y(x) = (B(x) \otimes B_Y^+) - (B(x) \otimes B_Y^-) \tag{8}$$

Four binary convolutions, i.e. $B(x) \otimes B_{X,Y}^{\pm}$, are used in $G_X(x)$ and $G_Y(x)$. $B(x) = \sum_{p=0}^{N-1} s(i_p, i_x) \cdot 2^p$ is the N-bit local binary pattern descriptor of a pixel $x$. $i_x$ is the central pixel intensity and $i_p$ is the intensity of the $p$-th neighbour of $x$ in the source pattern. The comparison operator is defined as:

$$s(i_p, i_x) = \begin{cases} 1 & if \; |i_p - i_x| > T_x \\ 0 & otherwise \end{cases} \tag{9}$$

where $T_x = \frac{1}{4} i_x + 20$ and $N = 5 \times 5 - 1$.

The range of the local binary pattern descriptor of a pixel is presented in gray color in Fig. 6(a). In Fig. 6(b), a "Square" source image is made-up of a solid black $10 \times 10$ pixel square on a $24 \times 24$ pixel white background. In the grayscale image, the intensities of white and black pixels are 255 and 0, respectively. For example, the intensity of the red-highlighted pixel $x$ in Fig. 6(b) is $i_x$ =255. We arrange and serialize the pixels in the range of local binary pattern descriptor by columns. The 1st neighbour pixel intensity is $i_1$ =0, hence according to Eq. (9), $s$ $(i_1, i_x)$=1; The 3rd neighbour is $i_3$ =255, hence, $s$ $(i_3, i_x)$=0. $B(x)$ can be therefore calculated for the red-highlighted pixel in Fig. 6(b) as follows:

$$B(x) = \begin{bmatrix} s(i_1, i_x) & - & - & - & s(i_{20}, i_x) \\ s(i_2, i_x) & - & - & - & s(i_{21}, i_x) \\ s(i_3, i_x) & - & x & - & s(i_{22}, i_x) \\ s(i_4, i_x) & - & - & - & s(i_{23}, i_x) \\ s(i_5, i_x) & - & - & - & s(i_{24}, i_x) \end{bmatrix} = \begin{bmatrix} 1 & 1 & 1 & 1 & 1 \\ 1 & 1 & 1 & 1 & 1 \\ 0 & 0 & x & 0 & 0 \\ 0 & 0 & 0 & 0 & 0 \\ 0 & 0 & 0 & 0 & 0 \end{bmatrix} \tag{10}$$

For the red-highlighted pixel $x$ in Fig. 6(b), "1" in $B(x)$ corresponds to a white pixel and "0" corresponds to a black pixel in the source image.

In Eqs. (7) and (8), $B_X^+$, $B_X^-$, $B_Y^+$ and $B_Y^-$ are the four kernels that are adopted as in Ref. [33]. Figure 6(c) shows the areas of the four different kernels. Pixels which fall outside the highlighted areas in Fig. 6(c) for a given string are set to zero. For example:

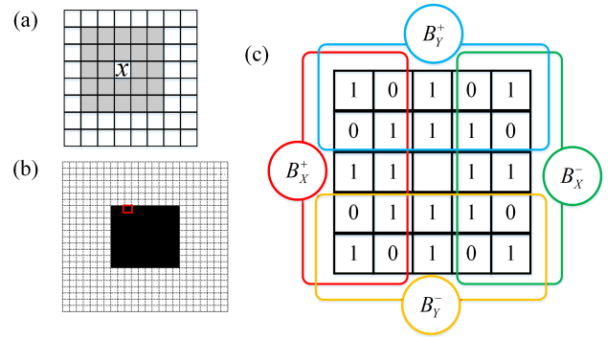

**Fig. 6** (a) Gray color: range of the local binary pattern descriptor of pixel $x$. (b) A $24 \times 24$ pixel source "Square" image. The red-highlight indicates a given pixel in the image. (c) The four convolutions $(B_X^+, B_X^-, B_Y^+$ and $B_Y^-)$ of the $5 \times 5$ binary pattern. Bits which fall outside the highlighted areas for a given string are set to zero.

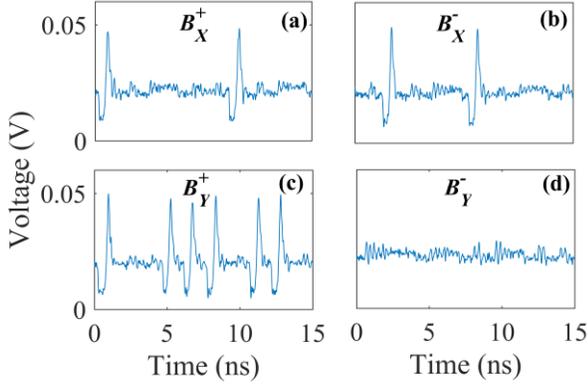

**Fig. 7.** Four convolutional results with four highlighted areas kernels for one pixel which has red box in Fig. 6.

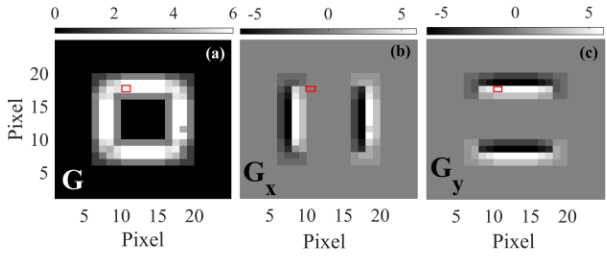

**Fig. 8.** Gradient maps of the "Square" source image. Visualizations of (a) $G$, (b) $G_X$ and (c) $G_Y$ maps of the "Square" source image based on the optical binary convolution performed by the VCSEL neuron.

$$B_X^+ = \begin{bmatrix} 1 & 0 & 0 & 0 & 0 \\ 0 & 1 & 0 & 0 & 0 \\ 1 & 1 & 0 & 0 & 0 \\ 0 & 1 & 0 & 0 & 0 \\ 1 & 0 & 0 & 0 & 0 \end{bmatrix}$$

(11)

We arrange and serialize the pixels of $B(x)$ and the four kernels by columns. For example, the string of $B(x)$ is [1, 1, 0, 0, 0, 1, 1, 0, 0, 0…] and the string of $B_x^+$ is [1, 0, 1, 0, 1, 0, 1, 1, 1, 0…]. We studied experimentally the response of the VCSEL neuron under the injection of the "Square" source image and kernel operators included in Figs. 6(b) and (c). Specifically, Fig. 7 showcases the experimentally-recorded results at the VCSEL output for each kernel when operating on the red-highlighted pixel in Fig. 6(b). It can be seen in Fig. 7(a) that fast (sub-ns) spikes are only triggered by the 1st and 7th pixels. Therefore, the convolutional result for $B(x) \otimes B_X^+$ is 2, as it was expected. Similarly, from Figs. 7(b) - (d) we can see that 2, 6 and 0 sub-ns spikes are elicited at the VCSEL's output for kernels $B_x^-$, $B_y^+$ and, $B_y^+$, respectively. Using the experimental results measured from the spiking VCSEL neuron we calculate off-line $G_X(x)$, $G_Y(x)$ and $G(x)$ to determine the image gradient magnitude. Based on the experimentally-measured results in Figs. 7(a) - (d), $G_X(x)$, $G_Y(x)$ and $G(x)$ are 0, 6 and 6 respectively using Eqs. (7) - (9).

The experimental process in Fig. 7 is repeated consecutively for every single pixel in the "Square" source image (Fig. 6(b)) to calculate their image gradient magnitudes. The latter are used to build the reconstructed image in Fig. 8(a), providing a gradient map for the "Square" source image. Figure 8(a) clearly reveals a "hollow" square shape in the experimentally produced gradient map; hence detecting all edge-features of the source image. In Fig. 8(a), the pixels with a gradient magnitude $G(x) > 3$ can be selected to thin the response and reveal the true edges of the "Square" [33, 34]. Additionally, Figs. 8(b) and 8(c) plot separately the reconstructed images using the obtained values for $G_X(x)$ and $G_Y(x)$ from the experimentally measured time-series at the VCSEL neuron's output. Figures 8(b) and 8(c) reveal that both vertical and horizontal lines can be individually detected from the source image in Fig. 6(b) using respectively the magnitudes $G_X(x)$ and $G_Y(x)$.

To further investigate our experimental system, we focused on demonstrating the achievement of gradient maps from a complex source image using the all-optical binary convolution of this work, as seen in Fig. 9. For this purpose, we selected as a source image for our VCSEL based binary convolution system a complex "Horse head" image (Fig.9(b)). This is a $100 \times 105$ pixel portion of the "Horse" image from the Berkeley Segmentation Data Set [32] (also included in Fig.9(a)). The colour image was converted to grayscale before we applied the same experimental methods used previously to obtain the results included in Fig. 8 above. The values of $G(x)$, $G_X(x)$ and $G_Y(x)$ experimentally achieved for the complex "horse head" image (Fig.9(b)) are shown in Figs. 9(c) - (e), respectively. These gradient maps reveal the successful detection of the edge features in this complex image; hence permitting the successful recreation of the outline and shape of the horse head. This effectively demonstrates that the reported all-optical binary convolution technique with a VCSEL neuron is also suitable for complex high-resolution source images.

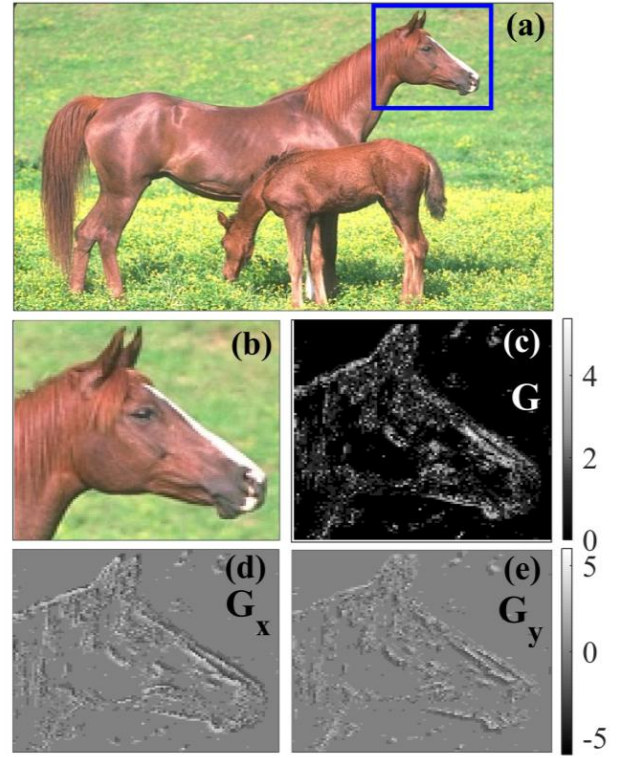

**Fig. 9.** The "Horse head" image and the gradient maps of the "Horse head" image. (a) Source "Horse" image. The blue box indicates the 'Horse Head' image used for analysis (b). Visualizations of the (c) $G$, (d) $G_X$ and (e) $G_Y$ maps of the "Horse head" image obtained from the optical binary convolution performed with the VCSEL neuron.

### C. NUMERICAL RESULTS

In this section, binary convolution based on a single VCSEL neuron is performed numerically. The robustness of the system to perform all-optical binary convolution under noisy inputs and for larger kernels is investigated. Finally, the calculation of image gradient magnitudes with our photonic approach using a single VCSEL neuron is presented numerically using the 'Horse' image from the latest version of the Berkeley Segmentation Data Set [32].

The binary convolution example given in Fig. 3 and experimentally performed with the VCSEL neuron (see Fig. 4) is numerically simulated using the SFM model in Figs. 10(a1) - (c1). Pixels of value "1" are numerically implemented using power drop pulses with strength $K_p$ = 0.852 ($K_p$ = pulse power/ constant power) with a duration of 1.5 ns (as in the experimental demonstration). The frequency detuning between the externally injected signal and the XP mode in the VCSEL model is set to -3.66 GHz. Figures 10 (a1) - (c1) plot the numerically obtained results for the all-optical binary convolution with a VCSEL neuron. Specifically, Figs. 10(a1) and (b1) plot respectively the time series for the temporally encoded image (Fig. 10(a1)) and kernel (Fig. 10(b1)) inputs, whilst Fig. 10(c1) plots the numerically calculated output from the VCSEL neuron. The latter clearly shows that the simulation successfully reproduces the outcome of the experimental all-optical binary convolution (see Fig 4(d)) where 4 spikes are elicited by the VCSEL. This excellent agreement between the modelled results and the experimental findings gives us confidence to test the robustness of the photonic binary convolution system under the injection of inputs with added noise. To study this aspect, we model the response of the VCSEL binary convolutional system under the injection of noisy inputs with a configured SNR= 20 dB (see results in Figs. 10(a2) and 10(b2)). Specifically, Fig. 10 (c2) shows that the exact same response is obtained from the VCSEL neuron as compared to the case with no added noise in Fig. 10(c1). This outlines the robustness to noise of the proposed all-optical VCSEL convolutional system. Additionally, the numerical convolution with a larger $5 \times 5$ pixel kernel is tested numerically in Figs. 10(a3) - 10(c3) using Eq. (10)

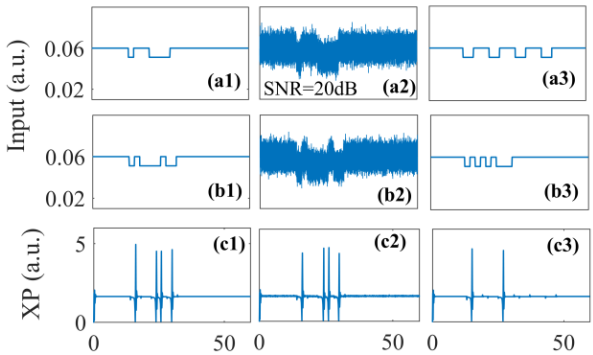

**Fig. 10.** (a1) - (a3) Inputs of channel 1 (Image in Fig. 3). (b1) - (b3) Inputs of channel 2 (Kernel in Fig. 3). (c1) - (c3) VCSEL neuron's output. (a1) - (c1) Convolutional operation in the VCSEL neuron without noise. (a2) - (c2) Convolutional operation in the VCSEL neuron with added inputs noise of SNR=20 dB. (a3) - (c3) Convolution operation with a $5 \times 5$ pixel kernel.

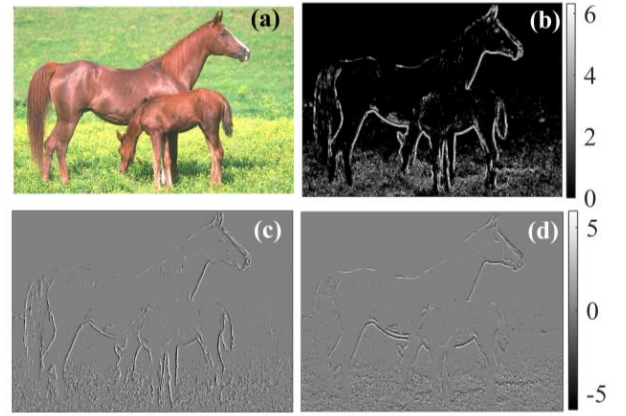

**Fig. 11** "Horse" image and gradient maps of the "Horse" image. (a) "Horse" image. Visualizations of $G$ (b), $G_X$ (c) and $G_Y$ (d) maps of "Horse" image based on the numerical optical binary convolution in VCSEL.

and Eq. (11) as inputs. Figure 10(c3) shows that the modelled convolutional result obtained from the VCSEL neuron also produces two fast spike events; hence yielding the exact same outcome as obtained experimentally in Fig. 7(a). We can therefore deduce that the convolution results that can be obtained with our VCSEL neuron based approach are not limited by the dimension of the kernel operators or the resolution of the image.

Figure 11 shows the numerically calculated gradient maps obtained with a spiking VCSEL neuron for the "Horse" source image [32] with a resolution of 481×321 pixels (Fig.11(a) and Fig. 9(a)). Figures 11(b-d) show the calculated gradients maps for $G(x)$, $G_X(x)$ and $G_Y(x)$, respectively. These were obtained using the $5 \times 5$ kernel introduced in the experimental study of the "Square" source image (see Figs. 6-8). It can be seen that the numerical simulation successfully reveals the image edge information through the gradient magnitude $G(x)$, as seen in Fig. 11(b), as well as the individual horizontal and vertical edge features of the source image through $G_X(x)$ and $G_Y(x)$, as seen in Figs. 11(c) and 11(d), respectively. These results, showing good overall agreement with the experimental findings of Fig. 9, therefore numerically validate that the gradient magnitude can be successfully calculated with a photonic spiking VCSEL-neuron, irrespectively of the image dimensionality.

### 4. CONCLUSION

In this work, we proposed and investigated experimentally and numerically an all-optical binary convolution system using a VCSEL operating as a photonic spiking neuron. The inputs (image and kernel) are encoded temporally using fast rectangular pulses (1.5 ns-long) and optically injected into the VCSEL neuron. The latter's optical output directly provides the results of the convolution in the number of (sub-ns long) spikes fired. In addition to performing all-optical binary convolution, we demonstrated experimentally and numerically the ability of the proposed system to calculate the image gradient magnitudes from digital source images. This feature was successfully used to identify key edge features from a source image as well as its separate horizontal and vertical components. Furthermore, we investigated numerically the robustness of the proposed VCSEL-based convolutional system to input noise. This simple system, using a single commercially-available VCSEL operating at the key telecom wavelength of 1300 nm, offers a novel photonic solution to binary convolution with the advantage of being highly energy efficient and hardware friendly.

This opens exciting prospects for a new photonic spiking platform for future optical binary spiking CNNs. Furthermore, the high-speed, low cost and neuronal functionalities of these photonic spiking systems hold promise for numerous processing tasks expanding into fields such as computer vision and artificial intelligence.

**Funding Information.** Office of Naval Research Global (ONRGNICOP-N62909-18-1-2027); the European Commission (828841-ChipAI-H2020-FETOPEN2018-2020); the UK's EPSRC Doctoral Training Partnership (EP/N509760). National Natural Science Foundation of China (61674119, 61974177); China Scholarship Council;

**Acknowledgments.** We thank Prof T. Ackemann and Prof. A. Kemp (University of Strathclyde) for lending some of the equipment used in this work.

**Disclosures.** The authors declare no conflicts of interest.